\documentclass[12pt]{article}
\usepackage{epsfig}
\usepackage{color}
\usepackage{amssymb,amsmath}
\usepackage{graphicx}
\usepackage{epsfig}

\newcommand{\bea}{\begin{eqnarray}}
\newcommand{\eea}{\end{eqnarray}}

 \makeatletter
\def\alt{\mathrel{\mathpalette\gl@align<}}
\def\agt{\mathrel{\mathpalette\gl@align>}}
\def\gl@align#1#2{\lower.6ex\vbox{\baselineskip\z@skip\lineskip\z@
\ialign{$\m@th#1\hfil##\hfil$\crcr#2\crcr\sim\crcr}}} \makeatother

\begin{document}
%
%
\vspace*{1.0cm}

\begin{center}
\baselineskip 20pt 
{\Large\bf 
Electroweak Phase Transition \\
in Georgi-Machacek Model
}
\vspace{1cm}

{\large 
Cheng-Wei Chiang$^{a,b,c}$, Toshifumi Yamada$^a$
} \vspace{.5cm}

{\baselineskip 20pt \it
$^a$ Department of Physics and Center for Mathematics and Theoretical Physics, \\
National Central University, Chungli, Taiwan 32001, Republic of China \\

$^b$ Institute of Physics, Academia Sinica, Taipei, Taiwan 11529, Republic of China \\

$^c$ Physics Division, National Center for Theoretical Sciences, Hsinchu, Taiwan 30013, Republic of China
}

\vspace{.5cm}

\vspace{1.5cm} {\bf Abstract} \end{center}

\ \ \ The Georgi-Machacek model extends the standard model Higgs sector by adding 
 two isospin triplet scalar fields and imposing global SU(2)$_R$ symmetry on them.
A feature of the model is that the triplets can acquire a large vacuum expectation value
 without conflicting with the current experimental bound on the $\rho$ parameter. 
We investigate the electroweak phase transition in the Georgi-Machacek model by evaluating the finite-temperature effective potential of the Higgs sector. 
The electroweak phase transition can be sufficiently strong in a large parameter space
 when the triplets acquire a vacuum expectation value of $O(10)$ GeV, opening a possibility to realize successful electroweak baryogenesis.

\thispagestyle{empty}

\newpage

\setcounter{footnote}{0}
\baselineskip 18pt
%

\ \ \ In spite of the discovery of a Higgs boson at the LHC \cite{higgs}, the structure of the Higgs sector remains largely unexplored.
The complete Higgs sector may not be the simplest one containing only one isospin doublet as in the standard model (SM), but may include multiple doublets, a singlet(s) or a triplet(s).
One of the motivations to consider such models with an extended Higgs sector comes from electroweak baryogenesis (EWBG) \cite{ewbg}, which is the only testable scenario for explaining the observed baryon asymmetry of the Universe.
It is well-known that successful EWBG relies on the following two conditions that are not met in the SM:
\begin{enumerate}
\item A strong first-order electroweak phase transition
 that enables decoupling of the sphaleron process in the symmetry broken phase so that created baryon asymmetry
 is not washed out.  More explicitly, the sphaleron process rate in the broken phase should be less than the Hubble parameter at that moment.
\item Large CP-violating phases that enable the creation of sufficient baryon asymmetry
 through scatterings off the bubble wall separating the broken and symmetric phases.
\end{enumerate}
Of the two conditions, the strong first-order phase transition is directly connected with
 the field content and structure of the Higgs sector,
 and hence motivates an extension of the Higgs sector beyond the SM.

Among various models with an extended Higgs sector, the Georgi-Machacek (GM) model with custodial vacuum alignment \cite{gm} has unique features that isospin triplet scalars can acquire a large vacuum expectation value (VEV), that this triplet VEV provides an origin for Majorana neutrino mass, and that the model predicts charged Higgs bosons to be tested at colliders \cite{cy,cky}.  
Due to a custodial symmetry that keeps the electroweak $\rho$ parameter unity at tree level, this model allows us to consider the scenario with a large triplet VEV.  In this letter, we study how a large triplet VEV of $O(1)$ GeV to $O(10)$ GeV affects the order of electroweak phase transition and whether a sufficiently strong first-order phase transition can be realized in some parameter space of the model.
In fact, since the Higgs potential of the GM model contains tree-level triple couplings involving two isospin doublet fields and one triplet or three triplet fields, we expect that a strong first-order phase transition can be achieved \textit{if} the triplets develop a large VEV and are as responsible for electroweak symmetry breaking (EWSB) as the doublet.

In our analysis of electroweak phase transition, we adopt a perturbative approach and 
 evaluate the finite-temperature effective potential at one-loop level \cite{dolanjackiw} for the GM model, from which the strength of phase transition, characterized by $v_C/T_C$, is evaluated, where $T_C$ and $v_C$ denote respectively the critical temperature and the scalar VEV at the critical temperature.
It is well-known that perturbation tends to break down at high temperatures \cite{parwani}, rendering the result of small $v_C/T_C$ less reliable, nevertheless it provides a good order-of-magnitude estimate of the phase transition strength.

This letter is organized as follows. We give quick reviews of the GM model, including its major theoretical and experimental constraints, and of the calculation of the one-loop finite-temperature effective potential.
We then conduct a numerical analysis on the strength of electroweak phase transition in various parameter regions of the GM model.
In particular, we identify two parameters on which $v_C/T_C$ has significant dependence and find regions with strong first-order phase transition.  Discussions of our findings are given toward the end of this letter, followed by a summary.
\\

The Higgs sector of the Georgi-Machacek (GM) model \cite{gm} contains 
 one isospin doublet scalar field with hypercharge $Y=1/2$, 
 denoted by $\phi=(\phi^+, \, \phi^0)^T=(\phi^+, \, \frac{1}{\sqrt{2}}(h_{\phi}+ia_{\phi}))^T$,
 one isospin triplet scalar field with $Y=1$, 
 denoted by $\chi=(\chi^{++}, \, \chi^+, \, \chi^0)^T=(\chi^{++}, \, \chi^+, \, \frac{1}{\sqrt{2}}(h_{\chi}+ia_{\chi}))^T$,
 and one isospin triplet scalar field with $Y=0$, 
 denoted by $\xi=(\xi^+, \, \xi^0, \, -(\xi^+)^*)^T=(\xi^+, \, h_{\xi}, \, -(\xi^+)^*)^T$.
Here $h_{\phi}, \, h_{\chi}, \, h_{\xi}$ are CP-even neutral components of the scalar bosons
 and $a_{\phi}, \, a_{\chi}$ are CP-odd ones.
On the Higgs potential, we impose a global SU(2)$_R$ symmetry that is explicitly broken by the SM Yukawa couplings.
To make the invariance under the SU(2)$_R$ transformation manifest, we write the Lagrangian in terms of the following SU(2)$_R$-covariant combinations of fields:
\begin{align}
\Phi \ &\equiv \ \left( \epsilon_2 \phi^*, \, \phi \right)
\ = \ \left(
\begin{array}{cc}
(\phi^0)^* & \phi^+ \\
-(\phi^+)^* & \phi^0
\end{array}
\right)
\ , \\
\Delta \ &\equiv \ \left( \epsilon_3 \chi^*, \, \xi, \, \chi \right)
\ = \ \left(
\begin{array}{ccc}
(\chi^0)^*    & \xi^+ & \chi^{++} \\
-(\chi^+)^*   & \xi^0 & \chi^+ \\
(\chi^{++})^* & -(\xi^+)^* & \chi^0
\end{array}
\right) 
\ ,
\end{align}
where
\begin{align}
\epsilon_2 = \left( \begin{array}{cc}
0 & 1 \\
-1 & 0
\end{array}
\right) 
~,~
\epsilon_3 \ = \ \left( \begin{array}{ccc}
0 & 0 & 1 \\
0 & -1 & 0 \\
1 & 0 & 0
\end{array}
\right)
~.
\end{align}
Under the SU(2)$_L$~$\times$~SU(2)$_R$ symmetry, $\Phi$ and $\Delta$ transform as $\Phi \rightarrow U_{2L} \Phi U^{\dagger}_{2R}$ and $\Delta \rightarrow U_{3L} \Delta U^{\dagger}_{3R}$,
 respectively,
 where $U_2$ is the two-dimensional representation of SU(2) group
 and $U_3$ is the three-dimensional one.
The Higgs sector Lagrangian is then expressed as
\begin{align}
{\cal L} \ =& \ 
\frac{1}{2} {\rm tr}[ (D^{\mu}\Phi)^{\dagger} D_{\mu}\Phi ]
\ + \ 
\frac{1}{2} {\rm tr}[ (D^{\mu}\Delta)^{\dagger} D_{\mu}\Delta ] 
\ - \ V_{tree}(\Phi, \, \Delta) \ - \ ({\rm Yukawa \ terms}) \ .
\end{align}
 where $D_{\mu}$ denotes the covariant derivative for $\Phi$ or $\Delta$.
Explicit expressions of the Yukawa terms, particularly those responsible for neutrino mass, can be found in Ref.~\cite{cy}, and are omitted here for simplicity.
The tree-level Higgs potential $V_{tree}(\Phi, \, \Delta)$ is given by
\begin{align}
V_{tree}(\Phi, \, \Delta) \ =& \ \frac{1}{2} m_1^2 \, {\rm tr}[ \Phi^{\dagger} \Phi ] + 
\frac{1}{2} m_2^2 \, {\rm tr}[ \Delta^{\dagger} \Delta ]
\nonumber \\
 &+ \lambda_1 \, \left( {\rm tr}[ \Phi^{\dagger} \Phi ] \right)^2 
 +  \lambda_2 \, \left( {\rm tr}[ \Delta^{\dagger} \Delta ] \right)^2 
 +  \lambda_3 \, {\rm tr}\left[ \left( \Delta^{\dagger} \Delta \right)^2 \right] 
 +  \lambda_4 \, {\rm tr}[ \Phi^{\dagger} \Phi ] {\rm tr}[ \Delta^{\dagger} \Delta ]
\nonumber \\
 &+  \lambda_5 \sum_{a,b=1,2,3} {\rm tr}\left[ \Phi^{\dagger} \frac{\sigma^a}{2} \Phi \frac{\sigma^b}{2} \right] 
                  {\rm tr}[ \Delta^{\dagger} T^a \Delta T^b]
\nonumber \\
 &+ \mu_1 \sum_{a,b=1,2,3} {\rm tr}\left[ \Phi^{\dagger} \frac{\sigma^a}{2} \Phi \frac{\sigma^b}{2} \right]
                               (P^{\dagger} \Delta P)_{ab}
 + \mu_2 \sum_{a,b=1,2,3} {\rm tr}[ \Delta^{\dagger} T^a \Delta T^b]
                               (P^{\dagger} \Delta P)_{ab} \ ,
\label{potential}
\end{align}
where $\sigma$'s are the Pauli matrices,
\begin{align}
&
T^1 = \frac{1}{\sqrt{2}} \left( \begin{array}{ccc}
0 & 1 & 0 \\
1 & 0 & 1 \\
0 & 1 & 0
\end{array}
\right)
~,~ 
T^2 = \frac{1}{\sqrt{2}} \left( \begin{array}{ccc}
0 & -i & 0 \\
i & 0 & -i \\
0 & i & 0
\end{array}
\right)
~,~ 
T^3 = \left( \begin{array}{ccc}
1 & 0 & 0 \\
0 & 0 & 0 \\
0 & 0 & -1
\end{array}
\right)
~,~ \nonumber \\
& \mbox{and } 
P = \frac{1}{\sqrt{2}} \left( \begin{array}{ccc}
-1 & i & 0 \\
0 & 0 & 1 \\
1 & i & 0
\end{array}
\right) \nonumber
~.
\end{align}
We take $m_1^2 < 0$ as in the SM and $m_2^2 > 0$.  In this case, the EWSB caused by the VEV of the doublet field will induce the triplet field to develop a VEV as well through the $\mu_1$ term.  Finally, we assume no CP violation in the newly introduced terms of the Lagrangian.

The EWSB vacuum at tree level is derived by solving the following tadpole conditions:
\begin{align}
\frac{\partial V(\Phi,\Delta)}{\partial h_{\phi}} \ &= \ \frac{\partial V(\Phi,\Delta)}{\partial h_{\chi}}
\ = \ \frac{\partial V(\Phi,\Delta)}{\partial h_{\xi}} \ = \ 0 ~,
\label{vacuum}
\end{align}
 with fields other than $h_{\phi},h_{\chi}$, and $h_{\xi}$ being zero.
From Eq.~(\ref{vacuum}), we choose the solution with the relation $h_{\chi}=\sqrt{2}h_{\xi}$, by which the EWSB vacuum maintains a diagonal SU(2)$_{L+R}$ or SU(2)$_V$ symmetry.
Writing the VEV's of $h_{\phi}, h_{\chi}, h_{\xi}$ as
 $\langle h_{\phi} \rangle = v_1, \ \langle h_{\chi} \rangle = \sqrt{2}v_2, \ \langle h_{\xi} \rangle = v_2$,
 respectively, we have 
 $\vert\langle h_{\phi} \rangle\vert^2 + 2\vert\langle h_{\chi} \rangle\vert^2 + 4\vert\langle h_{\xi} \rangle\vert^2
 =v^2\simeq(246$ GeV$)^2$.  
Here we define $\tan \theta_H$ as the VEV ratio, $\tan \theta_H \equiv 2\sqrt{2}v_{2}/v_1$.
When $v_1,v_2 \neq 0$,
 one can use Eq.~(\ref{vacuum}) to rewrite $m_1^2, m_2^2$ in terms of the VEV's of $h_{\phi}, h_{\chi}, h_{\xi}$ and other parameters in the Higgs potential as
\begin{align}
m_1^2 \ &= \ -4\lambda_1 v_1^2 - 6\lambda_4 v_2^2 - 3\lambda_5 v_2^2
- \frac{3}{2} \mu_1 v_2 \ , \nonumber
\\
m_2^2 \ &= \ -12\lambda_2 v_2^2 - 4\lambda_3 v_2^2 - 2\lambda_4 v_1^2
- \lambda_5 v_1^2 - \mu_1 \frac{v_1^2}{4v_2} - 6\mu_2 v_2 \ .
\label{m1m2}
\end{align}

We can derive the field-dependent mass matrices of the scalar bosons 
 from Eq.~(\ref{potential}) to be used in the evaluation of the finite-temperature effective potential.
After substituting $h_{\phi}=v_1$, $h_{\chi}=\sqrt{2}h_{\xi}=\sqrt{2}v_2$ and Eq.~(\ref{m1m2}) and diagonalizing the mass matrices, one finds three massless Nambu-Goldstone modes that eventually become the longitudinal components of the $W$ and $Z$ bosons.  
Under the classification of SU(2)$_V$ symmetry, the other massive states are grouped into
 a 5-plet $H_5 = (H_5^{++},H_5^{+},H_5^{0},H_5^{-},H_5^{--})^T$,
 a 3-plet $H_3=(H_3^{+},H_3^{0},H_3^{-})^T$, and two singlets $H_1^0$ and $H_1^{\prime0}$.  
Among these particles, only the 3-plet is CP-odd while the others are CP-even.  
The two singlets generally mix to produce physical states denoted by $H$ and $h$, where the latter is used to denote the recently discovered SM-like Higgs boson.  
As a result of the custodial symmetry, the components in each of the above-mentioned multiplets are degenerate in mass.  
Mass splittings of the order of a few hundred MeV due to electromagnetic breaking are expected within each representation, but can be safely ignored for our study.  
As a reference, we give their mass eigenvalues at the EWSB vacuum,
 with the help of Eq.~(\ref{m1m2}), as:
\begin{align}
\label{massH5}
m_{H_5}^2 \ =& \ 8\lambda_3 v_2^2 - \frac{3}{2}\lambda_5 v_1^2
- \frac{\mu_1 v_1^2}{4v_2} - 12 \mu_2 v_2 \ , \\
\label{massH3}
m_{H_3}^2 \ =& \ - \left(\frac{\lambda_5}2 
+ \frac{\mu_1}{4v_2} \right) v^2 \ , \\
\label{massH0}
m_{H,h}^2 \ =& \ 4\lambda_1 v_1^2 + 4(3\lambda_2+\lambda_3)v_2^2
- \frac{\mu_1v_1^2}{8v_2} + 3\mu_2 v_2 
\nonumber \\
& \pm \left\{ \, \left[4\lambda_1 v_1^2 + 4(3\lambda_2+\lambda_3)v_2^2
- \frac{\mu_1v_1^2}{8v_2} + 3\mu_2 v_2\right]^2 
- 4v_1^2v_2^2 \left[ 16\lambda_1(3\lambda_2+\lambda_3) - 3(2\lambda_4+\lambda_5)^2 \right]
 \right.
\nonumber \\
& \left.
+ 2\lambda_1\frac{\mu_1v_1^4}{v_2} + \frac{3}{4}\mu_1^2v_1^2
+ 6v_1^2v_2 ( 2\lambda_4 \mu_1 + \lambda_5 \mu_1 - 8\lambda_1 \mu_2) \right\}^{1/2} .
\end{align}

We comment on various limits of the GM model.
The triplet VEV $v_2$ vanishes when one sets $\mu_1=0$, as long as $m_2^2 > 0, \, \lambda_4 > 0$ and $\lambda_5 > 0$ are assumed.  In other words, the triplet VEV is induced by the doublet VEV through the $\mu_1$ term.
The GM model becomes SM-like, {\it i.e.}, the extra Higgs bosons $H_3, H_5, H$ are decoupled and the triplet VEV vanishes, when we take the limits of $\mu_1, v_2 \rightarrow 0$ with $\mu_1/v_2 \rightarrow \infty$.  A detailed discussion about the decoupling limit of the GM model is recently discussed in Ref.~\cite{logan}.
\\

Now we enumerate the theoretical and experimental constraints on the GM model 
 that will be incorporated in our parameter search for viable electroweak phase transition.
First, we consider the stability of the Higgs potential at large field values.  To avoid any runaway direction in the Higgs potential, 
 Refs.~\cite{arhrib,cy} have found the following conditions on the coupling constants $\lambda_i$:
\begin{align}
&
\lambda_1 > 0 ~, \ \lambda_2+\lambda_3 > 0 ~, \ \lambda_2 + \frac{1}{2}\lambda_3 > 0 ~,
\ -\vert \lambda_4 \vert + 2\sqrt{\lambda_1(\lambda_2+\lambda_3)} > 0 ~,
\nonumber \\
&
\lambda_4 - \frac{1}{4}\vert \lambda_5 \vert + \sqrt{2\lambda_1(2\lambda_2+\lambda_3)} > 0 ~.
\label{stability}
\end{align}

For perturbative calculations to be valid, we further impose the unitarity bound from the $S$-wave amplitudes for elastic scatterings of two scalar boson states.
The strongest bound as found in Ref.~\cite{ak} is:
\begin{align}
\vert 12 \lambda_1 + 22 \lambda_2 + 14 \lambda_3 \pm
\sqrt{(12\lambda_1 - 22\lambda_2 - 14\lambda_3)^2+144\lambda_4^2} \vert \ &< \ 16\pi \ .
\label{univio}
\end{align}

In subsequent analyses, we will restrict ourselves to the parameter space with $\tan \theta_H < 0.5$ ($v_2 < 39$ GeV).  This choice is made so that the constraints from both the $Z \rightarrow b \bar{b}$ decay \cite{cy} and the electroweak $S$ parameter are satisfied in the entire region \cite{cky}.
\\

At the one-loop level, the finite-temperature effective potential of the Higgs sector is given by \cite{dolanjackiw}
\begin{align}
V^1(\varphi; \, T)
=& \ V_B(\varphi) \ + \ V_0^1(\varphi, \, \mu_R) \nonumber \\
& + \ \frac{T^4}{2\pi^2} 
\left[ \ \sum_{i \in {\rm Bosons}} n_i I_B( \ m_i(\varphi)^2/T^2 \ ) \ + \ 
\sum_{i \in {\rm Fermions}} n_i I_F( \ m_j(\varphi)^2/T^2 \ ) \ \right] \ ,
\label{eff}
\end{align}
 where $T$ denotes the temperature, $\varphi$ collectively denotes the values of the fields, 
 $m_i(\varphi)$ denotes the field-dependent mass of mass eigenstate $i$, and
 $n_i$ counts its degrees of freedom.
$V_B(\varphi)$ is the same function as Eq.~(\ref{potential})
 with the parameters $m_1^2, m_2^2, \lambda_1, \lambda_2, \lambda_3, \lambda_4, \lambda_5, \mu_1$, $\mu_2$
 being replaced by the corresponding bare parameters that 
 should be fixed by nine renormalization conditions at zero temperature.
$V_0^1(\varphi, \, \mu_R)$ is the one-loop effective potential at zero temperature renormalized at the scale $\mu_R$,
 which is given by~\cite{cw}
\begin{align}
V_0^1(\varphi, \, \mu_R)
=& \frac{1}{64\pi^2} \left[
\sum_{i \in {\rm Bosons}} n_i \, (m_i^2(\varphi))^2 \, \left\{ \log\left( \frac{m_i^2(\varphi)}{\mu_R^2} \right) - C_i \right\} \ \right.
\nonumber \\
&~~~~~~~ \left. - \ 
\sum_{i \in {\rm Fermions}} n_i \, (m_i^2(\varphi))^2 \, \left\{ \log\left( \frac{m_i^2(\varphi)}{\mu_R^2} \right) - C_i \right\}
\right]
\label{zeroeff}
\end{align}
 where $\mu_R$ is the renormalization scale and $C_i$'s are constants that depend on the renormalization scheme.
The functions $I_B, \, I_F$ are defined as
\begin{align}
I_B(a^2) \ &= \ \int_0^{\infty} {\rm d}x \ x^2 \log\left[1-\exp(-\sqrt{x^2+a^2})\right] \ ,
\\
I_F(a^2) \ &= \ \int_0^{\infty} {\rm d}x \ x^2 \log\left[1+\exp(-\sqrt{x^2+a^2})\right] \ ,
\end{align}
 respectively, and interpolating functions are employed in numerical evaluations.

We adopt the Landau gauge in our calculations.
We include in `Bosons' of Eq.~(\ref{eff}) the $W$ and $Z$ bosons, the (field-dependent) mass eigenstates of Higgs bosons and the would-be Nambu-Goldstone modes.
In `Fermions', we include only the SM top quark neglecting the other SM matter fields.
\\

We evaluate the strength of electroweak phase transition characterized by $v_C/T_C$, the ratio of the Higgs VEV at the critical temperature and the critical temperature $T_C$, in a wide range of parameter space in the GM model.
We choose $\tan \theta_H, \ \lambda_1, \ \lambda_2, \ \lambda_3, \ \lambda_4, \ \lambda_5, \ \mu_1$ and $\mu_2$ as the independent parameters.
We then fix the value of $\lambda_1$ by the requirement that the mass of the lightest CP-even boson be 126 GeV, the mass of the currently observed Higgs boson.

Two of the nine renormalization conditions require that the zero-temperature one-loop effective potential, $V_B+V_0^1$, have a minimum at 
 $h_{\phi}=v \cos \theta_H, \, h_{\chi}=\frac{1}{2}v \sin \theta_H$.
The relation $h_{\xi}=\frac{1}{2\sqrt{2}}v \sin \theta_H$ automatically follows from these conditions due to SU(2)$_R$ symmetry of the potential.
The other seven renormalization conditions require the matching of components of the scalar boson mass matrices and three-point coupling constants evaluated from the tree-level potential Eq.~(\ref{potential}) and those evaluated from the zero-temperature one-loop effective potential, $V_B+V_0^1$.

We note in passing that,
 since the field-dependent mass eigenstates include the would-be Nambu-Goldstone modes 
 that become massless for 
 $h_{\phi}=v \cos \theta_H, \, h_{\chi}=\frac{1}{2}v \sin \theta_H, \, h_{\xi}=\frac{1}{2\sqrt{2}}v \sin \theta_H$,
 some of the renormalization conditions
 apparently contain $\log 0$ singularity.
In fact, terms containing $\log 0$ singularity vanish in dimensional regularization if they originate from the integral
 $\int {\rm d}^D p/(2\pi)^D \, p^{\alpha}$ with $\alpha\neq-2$ and $D=4-2\epsilon$. 
The integral $\int {\rm d}^D p/(2\pi)^D \, p^{-2}$ gives $\Gamma(\epsilon)/(4\pi)^D$, which is subtracted in the $\overline{{\rm MS}}$ scheme leaving no finite terms \cite{leibbrandt-capper}.

\begin{figure}[thbp]
 \begin{minipage}{0.5\hsize}
  \begin{center}
   \includegraphics[width=80mm]{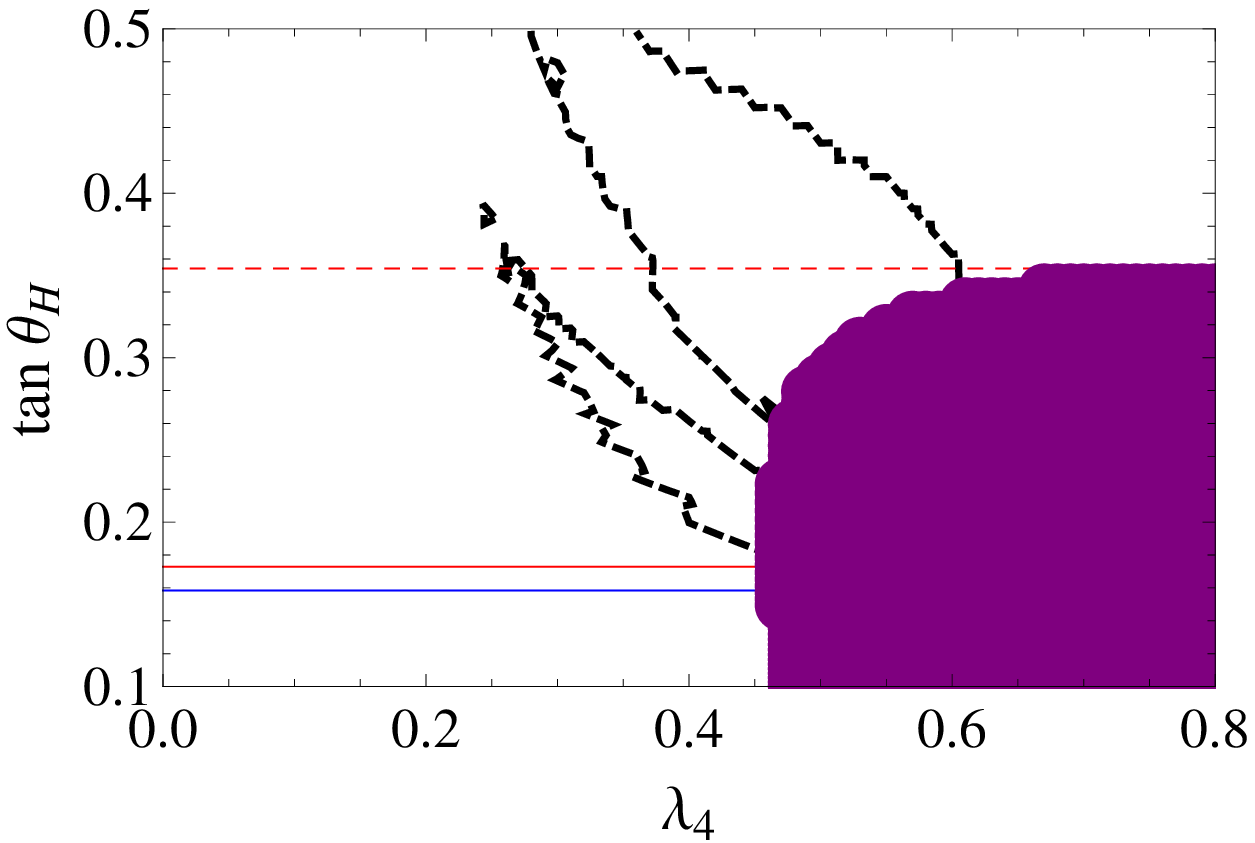}
  \end{center}
 \end{minipage}
  \begin{minipage}{0.5\hsize}
  \begin{center}
   \includegraphics[width=80mm]{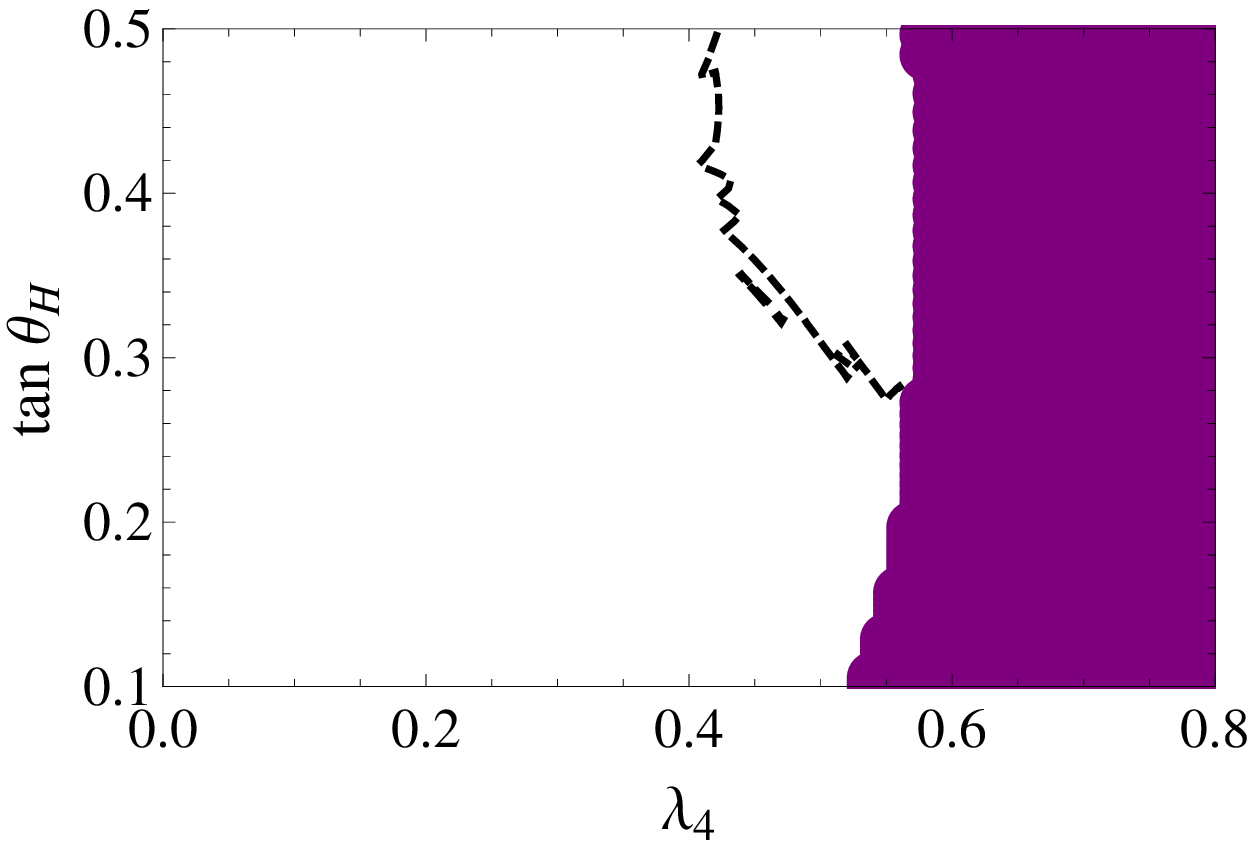}
  \end{center}
 \end{minipage}
 \caption{Contours of $v_C/T_C$ on the $\lambda_4$-$\tan \theta_H$ plane.  In both plots, we fix $\lambda_{2,3,5}=0.4$.  $\lambda_1$ is determined so that the lightest CP-even Higgs boson has the mass of 126 GeV.  The region filled by purple is unphysical due to vacuum instability.  
  The region surrounded by the thick black dashed curves corresponds to $1 < v_C/T_C < 2$, and that outside to $v_C/T_C < 1$.
  In the left plot, we take $\mu_1=\mu_2=-100$ GeV.  
  140~GeV$<m_{H_3}<$200~GeV above the red dashed line and 200~GeV$<m_{H_3}<$300~GeV between the red dashed and red solid lines.  200~GeV$<m_{H_5}<$300~GeV above the blue solid line.
  In the right plot, we take $\mu_1=\mu_2=-300$ GeV.
  Here $m_{H_{3,5}} >$ 300~GeV in the entire physically allowed region.
  \label{fig1}
  }
\end{figure}

\begin{figure}[htbp]
 \begin{minipage}{0.5\hsize}
  \begin{center}
   \includegraphics[width=80mm]{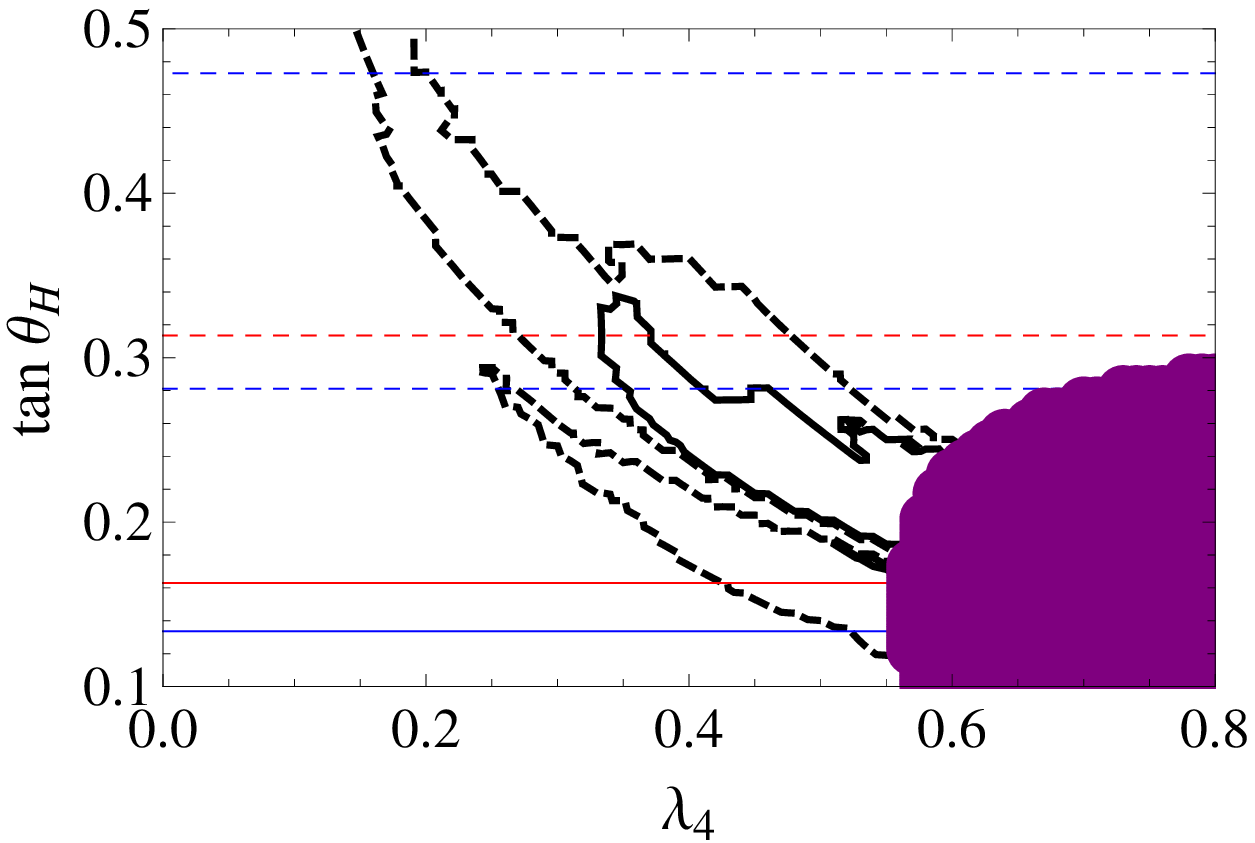}
  \end{center}
 \end{minipage}
  \begin{minipage}{0.5\hsize}
  \begin{center}
   \includegraphics[width=80mm]{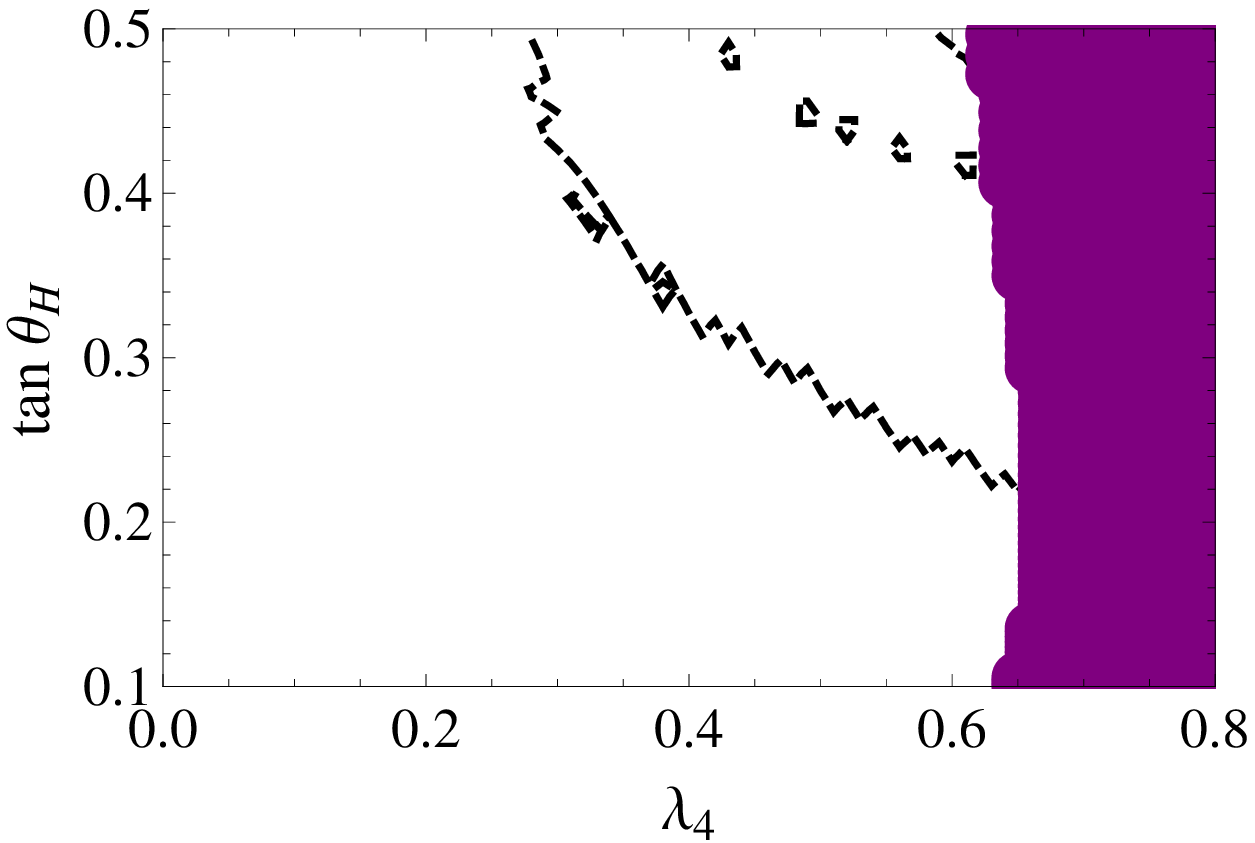}
  \end{center}
 \end{minipage}
 \caption{
 Same as Fig.~\ref{fig1}, but here we fix $\lambda_{2,3,5}=0.6$.  
 The region surrounded by the thick black solid curves corresponds to $2 < v_C/T_C < 3$.
 In the left plot, 140~GeV$<m_{H_3}<$200~GeV above 
  the red dashed line and 200~GeV$<m_{H_3}<$300~GeV between the red dashed and red solid lines.
 140~GeV$<m_{H_5}<$200~GeV between the blue dashed lines and
  200~GeV$<m_{H_5}<$300~GeV between the lower blue dashed and blue solid lines, and above the upper blue dashed line.
 In the right plot, both $m_{H_3}$ and $m_{H_5}$ are above 300~GeV in the entire physically allowed region.
  \label{fig2}}
\end{figure}

\begin{figure}[htbp]
 \begin{minipage}{0.5\hsize}
  \begin{center}
   \includegraphics[width=80mm]{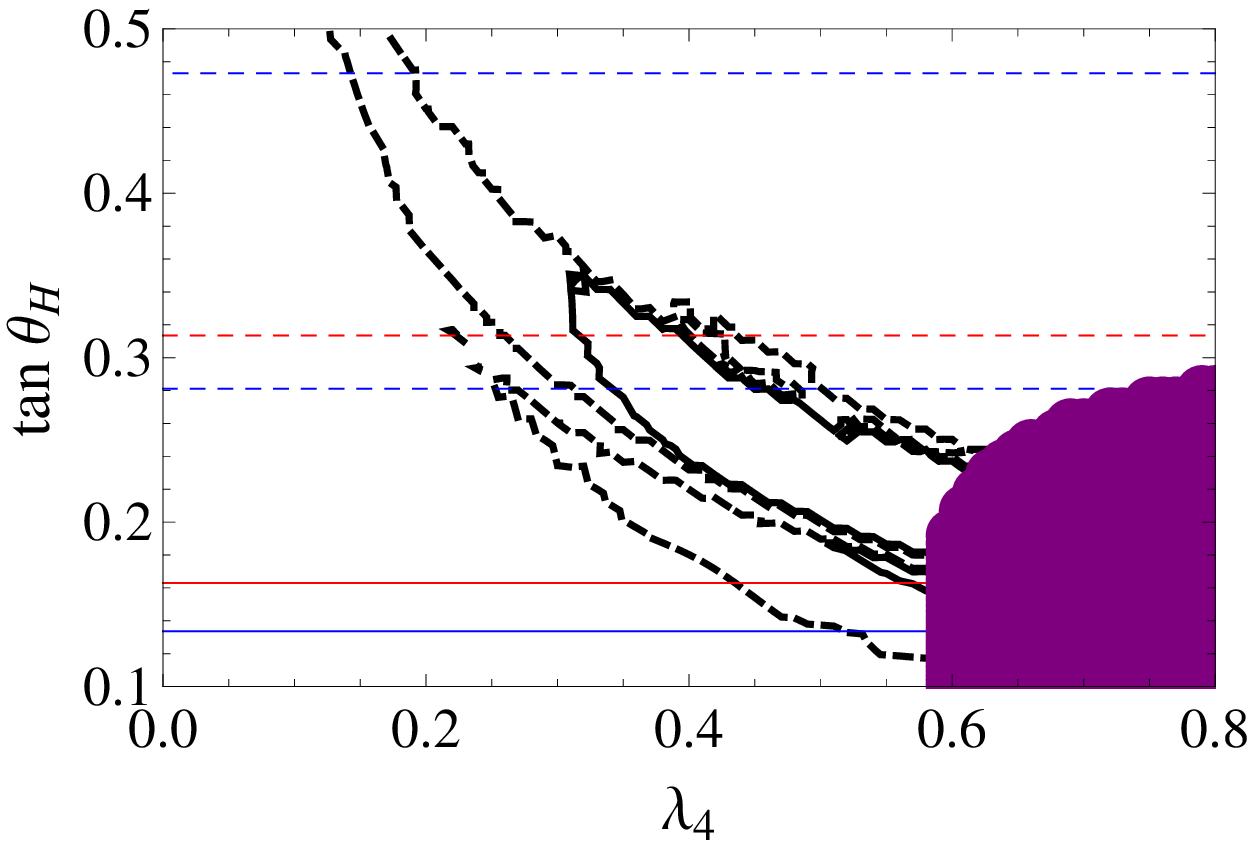}
  \end{center}
 \end{minipage}
  \begin{minipage}{0.5\hsize}
  \begin{center}
   \includegraphics[width=80mm]{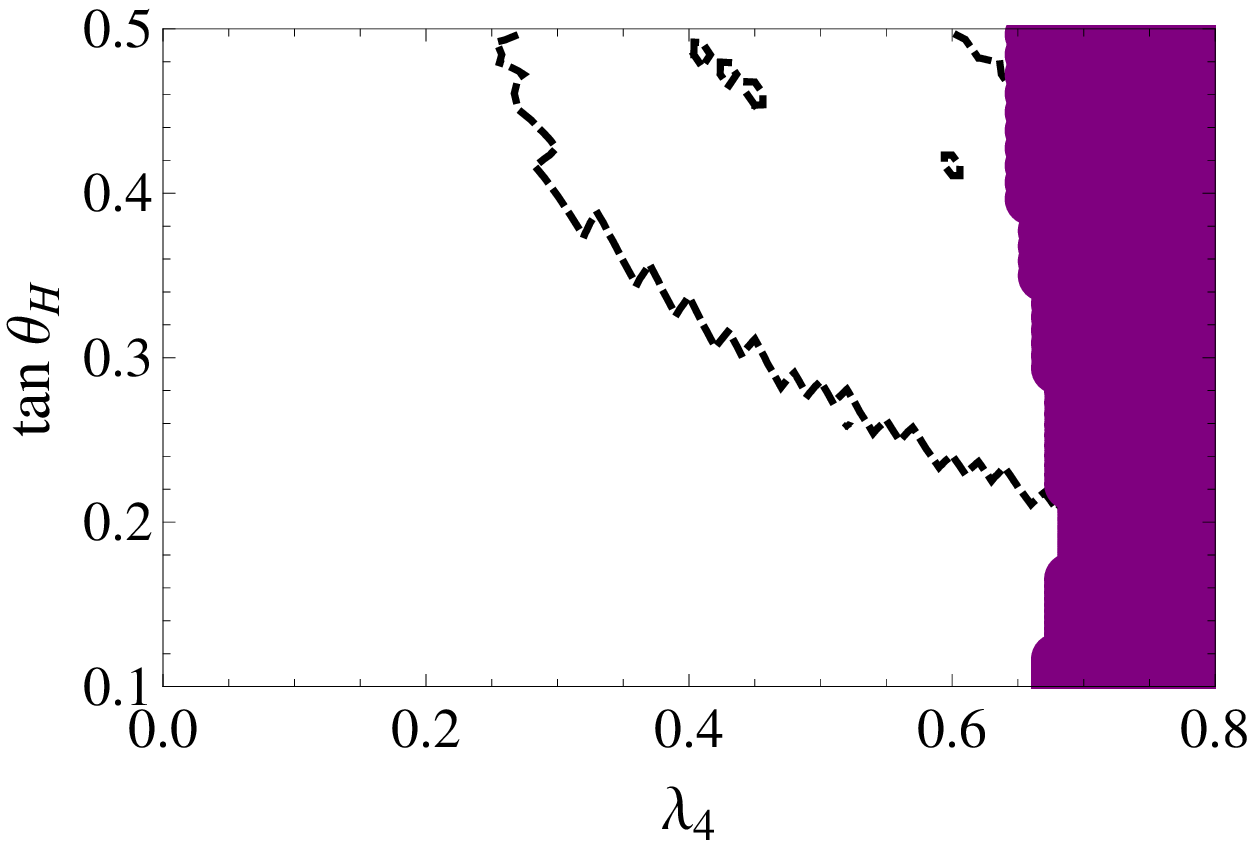}
  \end{center}
 \end{minipage}
 \caption{
 Same as Fig.~\ref{fig1}, but here we fix $\lambda_2=0.7,$ $\lambda_{3,5}=0.6$.
 Note that the domain of $2 < v_C/T_C < 3$ is slightly wider compared to the case with $\lambda_{2,3,5}=0.6$.
 \label{fig3}
 }
\end{figure}

After determining the bare parameters by the renormalization conditions,
 we numerically evaluate the critical temperature $T_C$ and the VEV of the fields at $T_C$, defined as
 $v_C \equiv \sqrt{ |\langle h_{\phi} \rangle_{T_C}|^2 + 2|\langle h_{\chi} \rangle_{T_C}|^2 
 + 4 |\langle h_{\xi} \rangle_{T_C}|^2 }$,
 using the finite-temperature one-loop effective potential in Eq.~(\ref{eff}).
In the course of numerical analysis,
 we exclude unphysical parameter regions where
\begin{itemize}
 \item the mass spectrum contains a negative squared mass;
 \item the potential is unbounded from below for large field values, 
 namely, the vacuum stability condition Eq.~(\ref{stability}) is not fulfilled;
 \item the perturbative unitarity condition Eq.~(\ref{univio}) is violated; or
 \item the electroweak symmetry breaking vacuum (with 
 $\langle h_{\phi} \rangle=v_1$, $\langle h_{\chi} \rangle=\sqrt{2}v_2$ and $\langle h_{\xi} \rangle=v_2$)
 is not the absolute minimum of the zero-temperature one-loop effective potential.
\end{itemize}
In our study, we have found that the strength of phase transition has a stronger dependence on $\lambda_4$ and $\tan \theta_H$ than on the other parameters in the Higgs potential.
In the following, we illustrate this by making contour plots of $v_C/T_C$ on the plane of $(\lambda_4, \, \tan \theta_H)$ while holding $\lambda_2, \lambda_3, \lambda_5, \mu_1$ and $\mu_2$ fixed.  
In Figs.~\ref{fig1} to \ref{fig3}, we pick three sets of parameters and focus on the region of $0 < \lambda_4 < 0.8$ and $0.1 < \tan \theta_H < 0.5$ to show the viability of strong first-order phase transition.
The condition on the strength of electroweak phase transition for successful baryogenesis is expressed as \cite{ewbg}
\begin{align}
v_C/T_C \ &\gtrsim \ \zeta, \label{sph}
\end{align}
where $\zeta$ is determined by the sphaleron decoupling condition and is usually about $1$.
In each plot, we enclose the region of $1< v_C/T_C < 2$ by the thick black dashed curves and that of $v_C/T_C > 2$ by the thick black solid curves.
We have found that, in the parameter space of interest, the vacuum stability condition Eq.~(\ref{stability}) provides the strongest constraint among those listed above,
 and hence we mark the regions where this condition is violated by the purple areas in the plots.

In Fig.~\ref{fig1}, we fix $\lambda_{2,3,5}=0.4$ and take $\mu_1=\mu_2=-100$ GeV in the left plot and $-300$ GeV in the right plot.
In Fig.~\ref{fig2}, we fix $\lambda_{2,3,5}=0.6$ with the same choices of $\mu_1, \mu_2$ for the left and right plots as in Fig.~\ref{fig1}.
In Fig.~\ref{fig3}, we fix $\lambda_2=0.7,$ $\lambda_{3,5}=0.6$, which are almost maximally allowed values, and $\mu_1$, $\mu_2$ are the same as in Fig.~\ref{fig1}.
If we take larger values for $\lambda_{2,3,5}$ ({\it e.g.}, $\sim 0.7$), then most of the parameter space is excluded 
 due to the violation of perturbative unitarity, Eq.~(\ref{univio}).
From Figs.~\ref{fig1}, \ref{fig2} and \ref{fig3}, we observe that strong first-order phase transition, $v_C/T_C \gtrsim 1$, can be achieved generally in the region with large $\tan \theta_H$ and large $\lambda_4$.  This is particularly apparent in the plots with $\mu_1=\mu_2=-300$ GeV. 
For the selected sets of parameters, the realization of $v_C/T_C \gtrsim 1$ requires $\tan \theta_H \gtrsim 0.15$, corresponding to $v_2 \gtrsim 13$~GeV.
This is consistent with our expectation that a large triplet VEV helps enhancing the strength of phase transition because it gives rise to a sizeable tree-level triple Higgs boson coupling.
We also find that the realization of strong first-order phase transition, since 
 the lower bound on $\tan \theta_H$ required for $v_C/T_C \gtrsim 1$ becomes smaller as the values of $\lambda_4$ and/or $\lambda_{2,3,5}$ increase.   
This is in accordance with the general argument that large quartic couplings for the Higgs boson and extra bosons enhance the value of $v_C/T_C$. In our case, radiative corrections from the additional bosonic fields to the finite-temperature effective potential are proportional to $\lambda_{2,3,4,5}$ and induce an effective triple Higgs coupling to trigger the first-order phase transition. 
However, if $-\mu_1=-\mu_2 = 100$~GeV, $v_C/T_C$ is below 1 for even larger values of $\tan \theta_H$ and $\lambda_4$.

The dependence of $v_C/T_C$ on $\tan \theta_H$ can be understood as follows.
Let us consider the following orthogonal transformation on the CP-even neutral fields in the GM model,
\begin{align}
\left(
\begin{array}{c}
h_1  \\
h_2  \\
h_3
\end{array}
\right)
\ &= \ 
A \begin{pmatrix}
0 & \frac1A\sqrt{\frac{1}{3}} & -\frac1A\sqrt{\frac{2}{3}} \\
\sqrt{3}\sin\theta_H & - \frac{4}{\sqrt{3}} \cos\theta_H & - 2\sqrt{\frac{2}{3}} \cos\theta_H \\
2\sqrt{2}\cos\theta_H & \sqrt{2}\sin\theta_H & \sin\theta_H
\end{pmatrix}
\left(
\begin{array}{c}
h_{\phi}  \\
h_{\chi}  \\
h_{\xi} 
\end{array}
\right)
\end{align}
 where $A\equiv1/\sqrt{8\cos^2\theta_H+3\sin^2\theta_H}$.
Among the fields on the left-hand side,
 only $h_3$ develops a VEV at zero temperature and hence plays a key role in the electroweak phase transition.
The field $h_3$ has a tree-level triple coupling that originates from the terms proportional to $\mu_1$ or $\mu_2$ in the Higgs potential
 (the last line of Eq.~(\ref{potential})), given by
\begin{align}
V_{tree}(\Phi, \, \Delta) \ &\supset \ 6A^3 \left( \mu_1 \cos^2\theta_H \sin\theta_H + \mu_2 \sin^3\theta_H \right) \, h_3^3 \ ,
\label{triple}
\end{align}
 independent of $\lambda_i$'s.
As $\tan\theta_H$ increases (but below 0.5), the first term on the right hand side of Eq.~(\ref{triple}) is enhanced and gives a significant contribution to the triple coupling of $h_3$.
Such a large triple coupling generally enhances the order of phase transition when $\mu_1$ is negative.
However, large $\tan \theta_H$ also induces a large quartic coupling of $h_3$ as 
\begin{align}
V_{tree}(\Phi, \, \Delta) \ &\supset
A^4 \left( 32 \lambda_1 \cos^4\theta_H + 9 \lambda_2 \sin^4\theta_H + 3 \lambda_3 \sin^4\theta_H
\right.
\nonumber \\
& \qquad~~~ \left. + 24 \lambda_4 \cos^2\theta_H \sin^2\theta_H + 12 \lambda_5 \cos^2\theta_H \sin^2\theta_H \right) \, h_3^4 \ .
\label{quartic}
\end{align}
The term proportional to $\lambda_4$, in particular, has a sizeable contribution if $\tan \theta_H$ and $\lambda_4$ are both large.
Such a large quartic coupling of $h_3$ in turn suppresses the order of phase transition, competing with the enhancement of the order of phase transition by the triple coupling.
Therefore, we observe that $v_C/T_C$ has the tendency of first increasing with $\tan \theta_H$ and then decreasing for even larger $\tan \theta_H$ when the other parameters are fixed.

\begin{figure}[htbp]
 \begin{minipage}{0.5\hsize}
  \begin{center}
   \includegraphics[width=80mm]{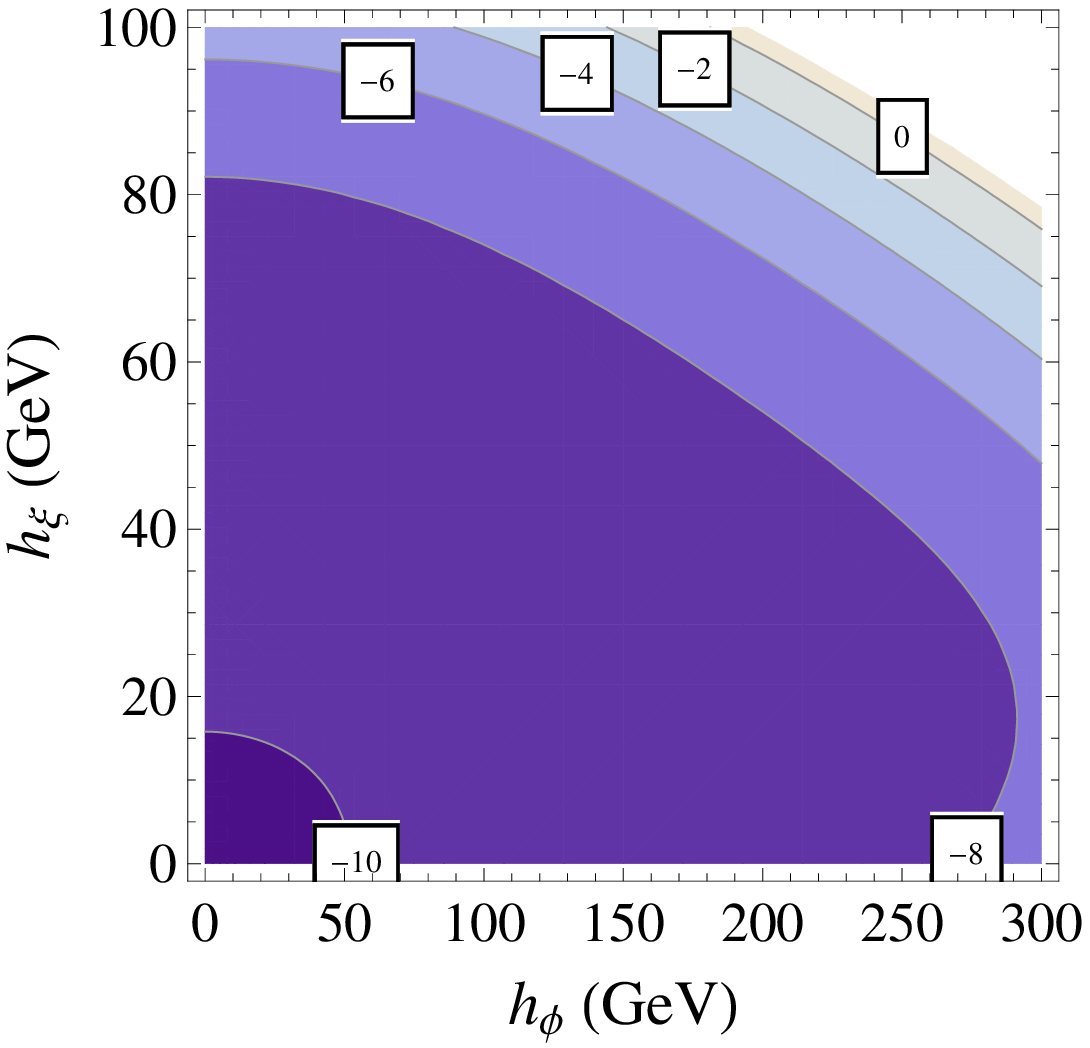}
  \end{center}
 \end{minipage}
  \begin{minipage}{0.5\hsize}
  \begin{center}
   \includegraphics[width=80mm]{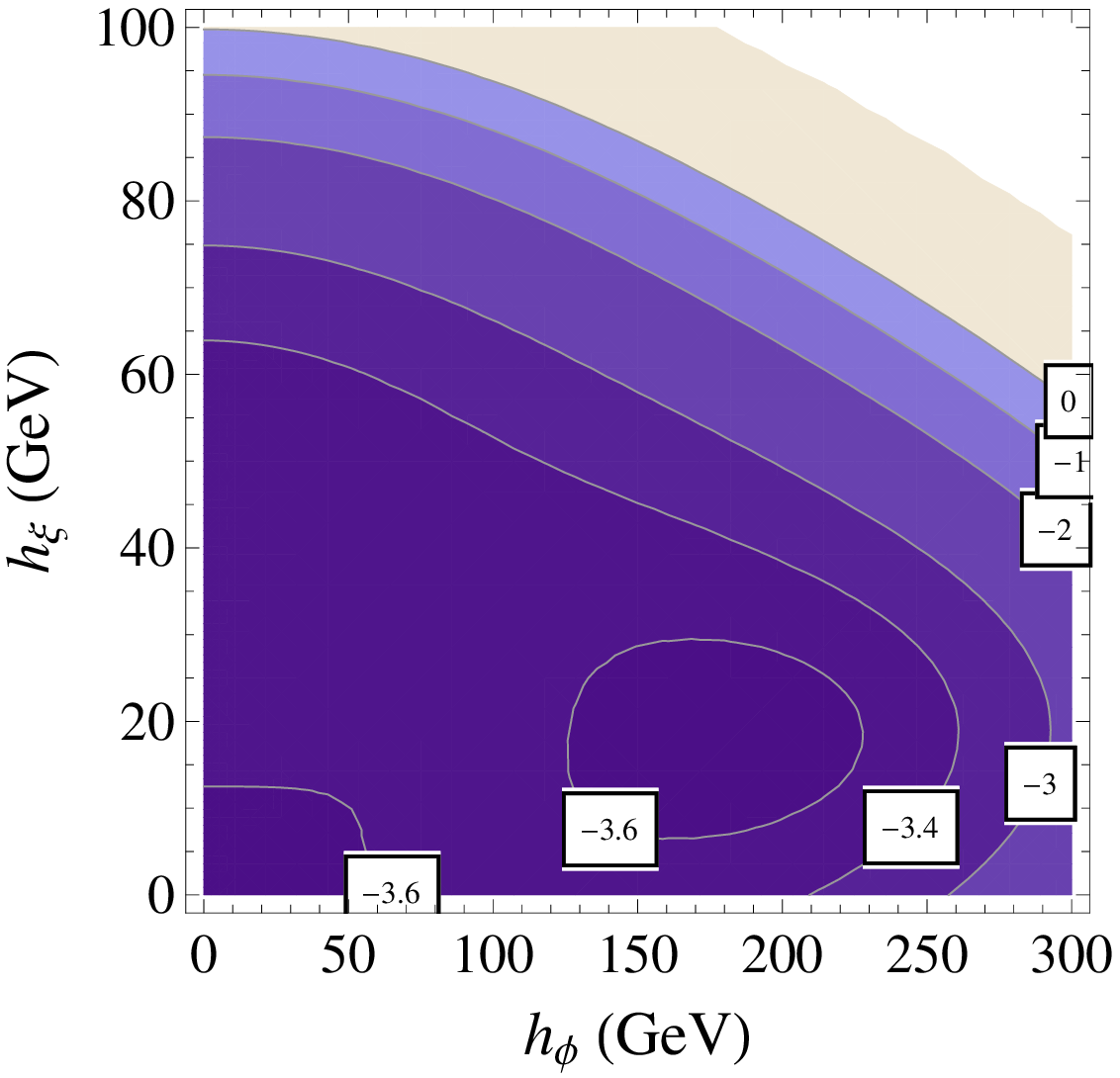}
  \end{center}
 \end{minipage}
 \begin{minipage}{0.5\hsize}
  \begin{center}
   \includegraphics[width=80mm]{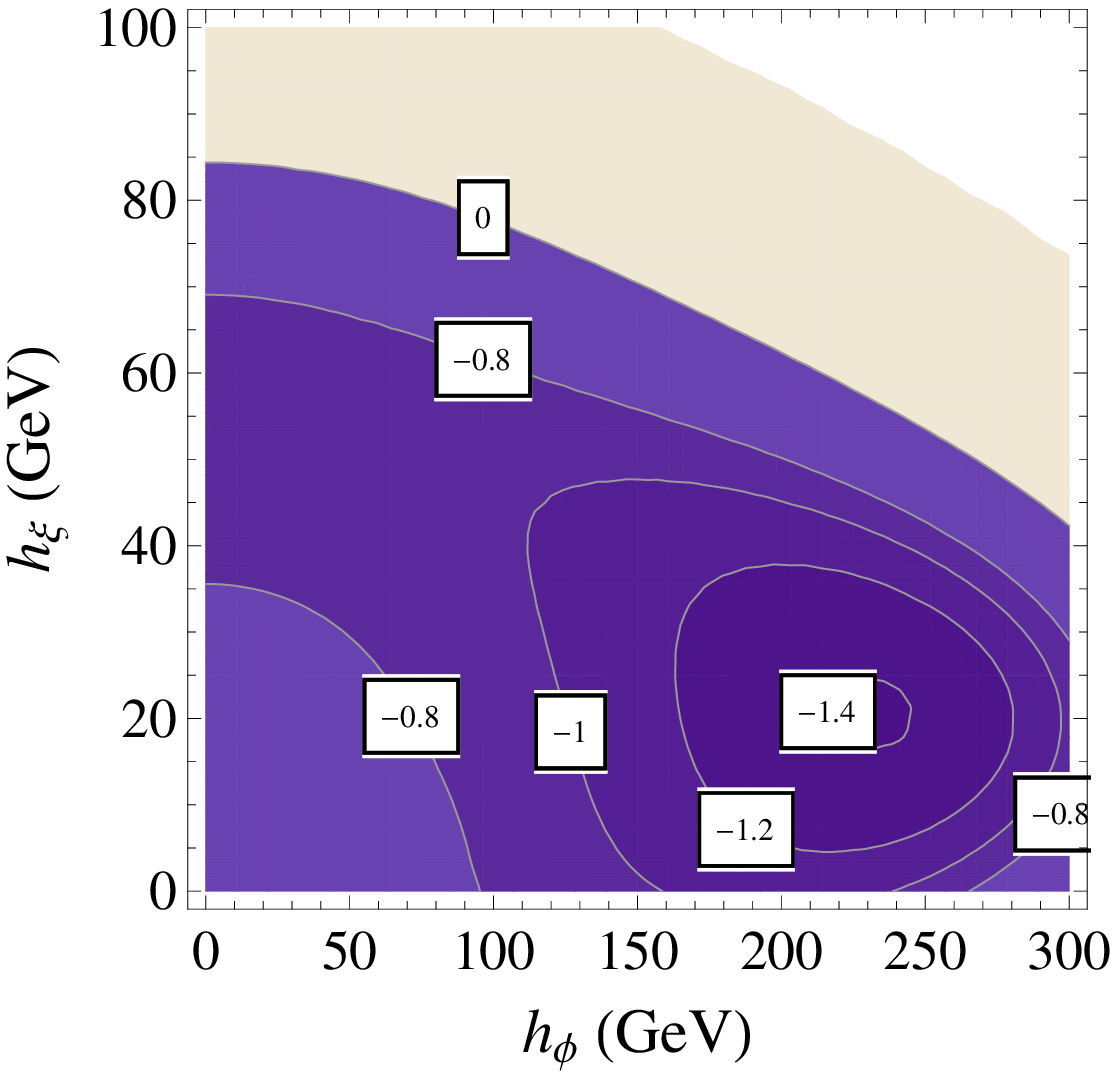}
  \end{center}
 \end{minipage}
  \begin{minipage}{0.5\hsize}
  \begin{center}
   \includegraphics[width=80mm]{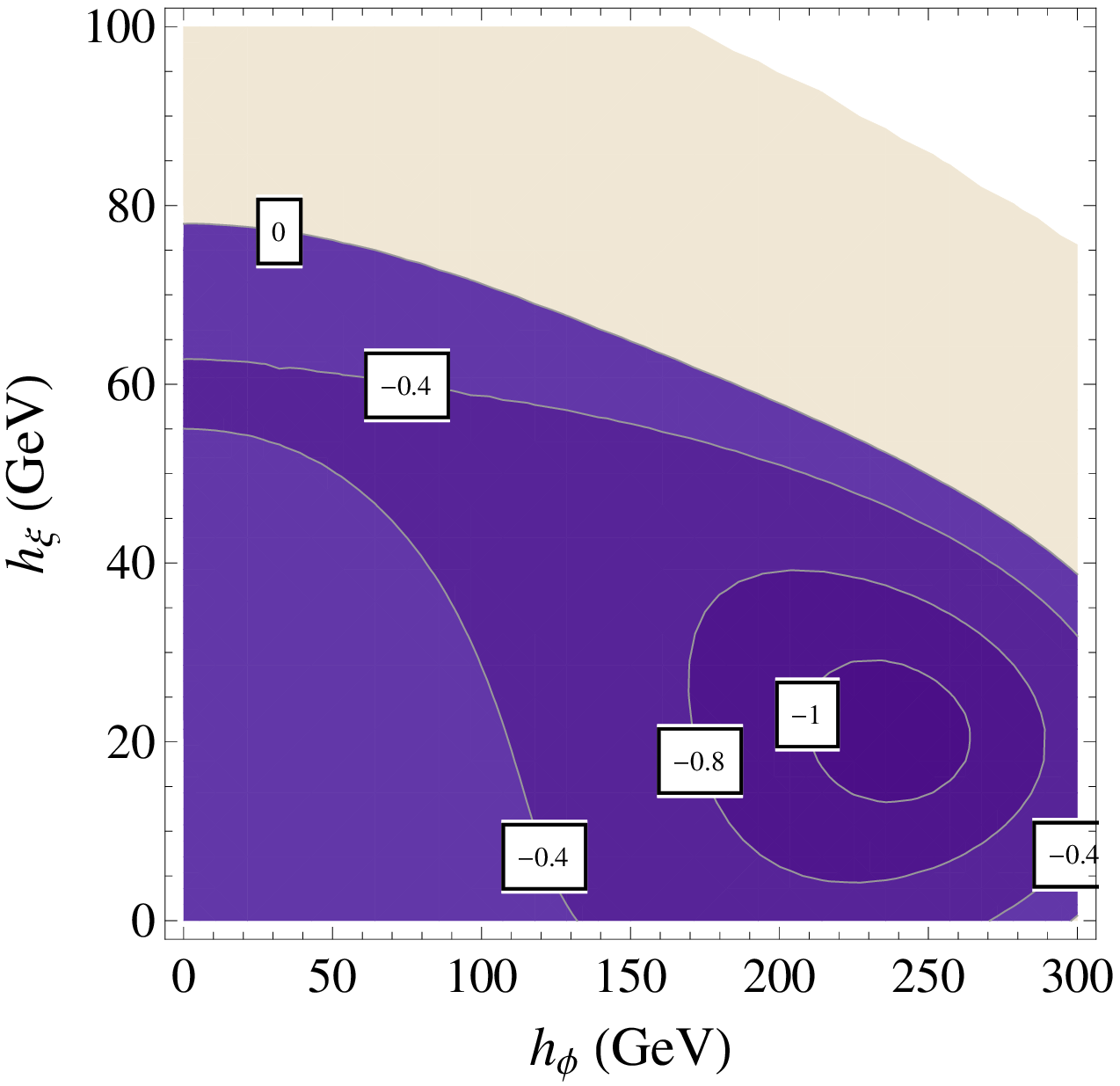}
  \end{center}
 \end{minipage}
 \caption{Contour plots of the value of the Higgs potential 
  at $T=120$~GeV (upper left), $T=92.5$~GeV $\simeq T_C$ (upper right), $T=60$~GeV (lower left), and $T=0$~GeV (lower right), 
  at the parameter point of $\lambda_{2,3,5}=0.6$, $\lambda_4=0.4$, $\mu_1=\mu_2=-100$~GeV and $\tan \theta_H=0.25$.
 The numbers labeled on the contours are in units of $(100$~GeV$)^4$.
 \label{contours}
 }
\end{figure}

To examine how such strong first-order electroweak phase transition occurs in the regions with $v_C/T_C > 2$, we take as one example the parameter choice of $\lambda_{2,3,5}=0.6$, $\lambda_4=0.4$, $\mu_1=\mu_2=-100$~GeV and $\tan \theta_H=0.25$, which sits right inside the region surrounded by the solid curve in Fig.~\ref{fig2}.
Fig.~\ref{contours} presents contour plots of the Higgs potential on the plane of $(h_{\phi}, \, h_{\xi})$ with $h_{\chi}=\sqrt{2}h_{\xi}$ assumed, at $T=120$~GeV, $T=92.5$~GeV $\simeq T_C$, $T=60$~GeV, $T=0$~GeV, where $T_C$ denotes the critical temperature for this parameter choice.
The number on each contour gives the value of the Higgs potential in units of $(100$~GeV$)^4$.
Here we observe that, as temperature approaches the critical temperature from above, a local minimum other than the origin develops and eventually becomes the absolute minimum below the critical temperature.

Collider phenomenology of the new Higgs bosons in the GM model at LHC had been extensively classified and analyzed in Ref.~\cite{cy}.  In particular, specific channels and kinematic cuts were proposed to search for such bosons.
Here we compute the masses of the SU(2)$_V$ 3-plet and 5-plet, $m_3$ and $m_5$, and superimpose their contours in Figs.~\ref{fig1} to \ref{fig3}.  As apparent from Eqs.~(\ref{massH5}) and (\ref{massH3}), $m_3$ and $m_5$ do not have any dependence on $\lambda_4$.
It turns out that both $m_3$ and $m_5$ are above 300~GeV in the entire physical region of each right plot of Figs.~\ref{fig1} to \ref{fig3}, where $\mu_1=\mu_2=-300$ GeV.
On the other hand, $m_3$ and $m_5$ can go below 300~GeV in the left plots, where $\mu_1=\mu_2=-100$ GeV. 
Note that since $m_3$ and $m_5$ are both above 140~GeV in the entire physically allowed domains in these figures, the mass spectra considered here safely evade the LEP bound on charged scalars.
\\

In summary, we have found the regions on the $\lambda_4$-$\tan\theta_H$ plane that grant strong first-order phase transition after fixing the other five parameters.  
If new sources of CP violation beyond the SM Yukawa couplings are provided, successful electroweak baryogenesis will be possible in these regions.
However, since all the coupling constants in the GM model are real if SU(2)$_R$ symmetry is imposed, and since terms that explicitly break SU(2)$_R$ symmetry are constrained by the experimental value of the $\rho$ parameter, it is necessary to further extend the GM model to incorporate additional CP-violating phases.
We have discovered that strong first-order phase transition characterized by $v_C/T_C \gtrsim 1$ is generally realized for large values of $\tan \theta_H$ and/or large values of $\lambda_4$ in the GM model.  The strength is further enhanced for larger quartic couplings, subject to the constraints of perturbative unitarity and Higgs potential stability.
For the selected parameter sets, the minimum of $\tan \theta_H$ with which strong first-order phase transition is viable is about $0.15$, corresponding to the triplet VEV $v_2 \simeq 13$~GeV.
\\

\section*{Acknowledgments}

C.-W.~C. thanks the hospitality of the Theoretical Particle Physics Group at Nagoya University during his visit and where part of this work was carried out.  T.~Y.~thanks Dr.~Kei Yagyu for useful discussions of the Georgi-Machacek model.  This research was supported in part by the National Science Council of R.O.C. under Grant Nos.~NSC-100-2628-M-008-003-MY4 and NSC-102-2811-M-008-019.

\end{document}